\documentclass[dvipsnames]{article}
\usepackage{adjustbox}
\usepackage{amssymb}
\usepackage{amsthm}
\usepackage{amsmath}
\usepackage{authblk}
\usepackage{blkarray}
\usepackage{booktabs}
\usepackage{bm}
\usepackage{calc}
\usepackage{caption}
\usepackage{cases}
\usepackage{centernot}
\usepackage[utf8]{inputenc}
\usepackage{dcolumn}
\usepackage{float}
\usepackage[margin=1in]{geometry}
\usepackage{graphicx}
\usepackage{listings}
\usepackage{makecell}
\usepackage{mathtools}
\usepackage{multicol}
\usepackage{pifont}

\usepackage{setspace}
\usepackage{soul}
\usepackage{subcaption}
\usepackage{tabularx}
\usepackage{tgpagella}
\usepackage{tikz}
    \usetikzlibrary{positioning}
\usepackage{tikzsymbols}
\usepackage{tkz-graph}
\usepackage{verbatim}
\usepackage{xcolor}
\usepackage{xfrac}
\usepackage{lineno}
\usepackage[style=english]{csquotes}
    \MakeOuterQuote{"}

\makeatletter
\def\bbordermatrix#1{\begingroup \m@th
  \@tempdima 4.75\p@
  \setbox\z@\vbox{%
    \def\cr{\crcr\noalign{\kern2\p@\global\let\cr\endline}}%
    \ialign{$##$\hfil\kern2\p@\kern\@tempdima&\thinspace\hfil$##$\hfil
      &&\quad\hfil$##$\hfil\crcr
      \omit\strut\hfil\crcr\noalign{\kern-\baselineskip}%
      #1\crcr\omit\strut\cr}}%
  \setbox\tw@\vbox{\unvcopy\z@\global\setbox\@ne\lastbox}%
  \setbox\tw@\hbox{\unhbox\@ne\unskip\global\setbox\@ne\lastbox}%
  \setbox\tw@\hbox{$\kern\wd\@ne\kern-\@tempdima\left[\kern-\wd\@ne
    \global\setbox\@ne\vbox{\box\@ne\kern2\p@}%
    \vcenter{\kern-\ht\@ne\unvbox\z@\kern-\baselineskip}\,\right]$}%
  \null\;\vbox{\kern\ht\@ne\box\tw@}\endgroup}
\makeatother

\newcommand{\col}{_\text{col}}

\newcommand{\row}{_\text{row}}

\newcommand{\conv}{\text{conv}}

\theoremstyle{definition}

\title{Worst-case control via linear programming: applications to truncation selection and partially malicious players}
\author[1]{Zhuoer Zhang \thanks{17zz13@queensu.ca}}
\author[2]{Bryce Morsky \thanks{bmorsky@fsu.edu}}
\affil[1]{Department of Physics, Queen's University, Kingston, ON, CA}
\affil[2]{Department of Mathematics, Florida State University, Tallahassee, FL, USA}
\date{\today}

\begin{document}

\maketitle

\begin{abstract}
    The connection between game theory, convex optimization, and geometry is deep. There are many applications of linear programming methods and polyhedral representation conversion methods in game theory. In this paper, we discuss two more scenarios where such methods can be useful. The first scenario is predicting the results of independent truncation dynamics under the large population assumption. The second scenario is when a player's opponent in a normal form game is not completely rational but shows some degree of malice. We show how one can compute a more profitable defensive play compared to simply playing a maximin strategy. We provide detailed computation procedure and numerical results for both scenarios.
\end{abstract}

\textbf{Keywords:} irrational players, linear programming, polyhedral representation conversion, truncation selections, worst-case control

\section{Introduction}

Game theory plays an important role in mathematically modeling behaviour and decision-making of players whether or not they are human \cite{myerson91}. This feature allows the theory to have applications in fields such as social sciences, economics, biology, and computer science. In particular, many studies have focused on two-player games, which are an essential foundation of many models of interactions between individuals. As the size of players' strategy sets increases, however, analyzing(say, finding the Nash equilibria) the multi-variable polynomial payoff expressions becomes computationally complex \cite{etessami10,papadimitriou07}. In such situations, support enumeration is a standard approach to finding Nash equilibria \cite{dickhaut93,porter08,vonstengel21}. Linear programming can be used as an auxiliary tool for support enumeration, and has been well applied to game theory \cite{lucchetti06, parrilo06,thie11}, often to compute minimax/maximin strategies, Nash equilibria, correlated equilibria, and related solution concepts. In this study, we convert problems into linear programs (or, more fundamentally, polyhedral conversion problems) and apply them to two additional scenarios: truncation selection and opponents that show a degree of irrationality or malice. We focus on these two scenarios because they both require consideration of worst-case scenarios, and they can be naturally expressed in linear algebraic language and converted to linear programs.

Before addressing irrational or malicious opponents, we begin with a discussion of rationality. In classical game theory, particularly for finite normal-form games, stable outcomes of games are predicted by Nash equilibria \cite{nash50} (and its variants), which rely on at least two assumptions: that payoff matrices accurately and quantitatively reflect what matters to each player, and that all players aim to maximize their payoffs. In practice, however, players cannot always assume that their opponents are rational and must instead make decisions based on payoffs for all possible scenarios. For example, players may encounter opponents that are only partially rationality \cite{htun05}, unfamiliar with the game \cite{burtonchellew16}, prone to making errors \cite{selten88}, or malicious \cite{moscibroda09}. Another typical reason for the assumption of rationality to fail is, in practice, payoff matrices usually only reflect the material payoffs of players, failing to capture players' complicated underlying emotions \cite{su24}. In the case of maliciousness, players may attempt to decrease their opponent’s payoff on purpose rather than maximize their own payoff. Such behaviour can be explained by social norms that promote costly punishment \cite{henrich06}, a component of relational utility in a player's payoffs \cite{su24}, or the game being a sub-game of a larger game where opponents' payoffs negatively affect one another. Players may also encounter games with incomplete information, such as Bayesian games \cite{harsanyi68}, where analyzing an opponent’s behaviour with certainty is impossible. In such situations, players may wish to guarantee themselves a minimum payoffs \cite{bruyere14} --- i.e.\, control for worst-case scenarios \cite{dave24}.

Risk averse or security strategies to handle such scenarios have been well studied \cite{binmore92,holler90,holler92,holler19}. A classic example is the maximin strategy, where a player assumes the worst-case scenario will happen and chooses a strategy that maximizes their payoff under that assumption. However, players may seek a balance between securing a lower-bound payoff and exploiting an opponent’s partial rationality. A safe equilibrium formalizes this trade-off by each player assigning probabilities over maximin and Nash equilibrium strategies \cite{ganzfried23}. When a player is willing to accept a worst-case payoff below the maximin value, they can respond more flexibly to educated beliefs about opponents' behaviours. This flexibility may lead to higher payoffs, particularly when combined with methods in behavioural game theory \cite{drew19,wright14}. In this paper, we design a model to capture the behaviour of partially malicious players: players who aim to increase their payoffs when they are low, but become malicious once they exceed a certain payoff threshold. While a fully malicious player reduces the problem to a classic zero-sum game, our framework generalizes this case to opponents exhibiting partial malice. We demonstrate how a rational player can enhance their payoff from the maximin payoff if the opponent follows such a decision-rule, and how such players would compute their ideal strategies. We analyze strategies for both rational players facing partially malicious opponents and malicious players seeking to minimize their opponents' payoffs under this decision-rule.

The population-level dynamics of avoiding worst-case outcomes can be modeled by truncation selection, an evolutionary algorithm in which only players with fitnesses (i.e.\ average payoffs) above a specified threshold survive and reproduce \cite{ficici06,ficici05,ficici00,ficici07,fogel97,fogel98,fogel11,jebari13,morsky16,morsky19}. Thresholds may be fixed or dynamic and can depend on variables such as time, which is particularly relevant for reproduction timing \cite{morsky20}. Truncation selection emphasizes adequacy: as long as a player's fitness exceeds the threshold, they propagate at an equal rate. This approach has applications in computer science and mathematical biology, including genetic algorithms \cite{jebari13} and as an alternative to the replicator equation \cite{morsky16}. Truncation selection can exhibit markedly different dynamics than the replicator equation \cite{fogel97}, where population change depends on deviations from the average fitness \cite{cressman14}. Truncation selection can be classified into two different methods: independent truncation selection, where the threshold is a fixed value; and dependent truncation selection, where a fixed proportion of the population survives, making the threshold contingent upon the population's fitness distribution \cite{morsky16,morsky19}. In this paper, we emphasize the role of linear programming and the concept of safe strategies in analyzing independent truncation, since the threshold condition can be naturally expressed by linear program constraints. In reformulating independent truncation selection as linear programs and polyhedral conversion problems, we offer a computationally efficient and analytically transparent framework for studying it.

\section{Methods}

\subsection{Risk aversion and partially malicious players}

We begin by defining some basic game theoretic concepts in linear algebraic terms, since this is a useful formulation for linear and polyhedron conversion problems (which are detailed in Appendices \ref{app:LP_methods} and \ref{app:PRC_methods}). A finite two-player normal-form game can be represented by a payoff table where the rows and columns represent the players' pure strategies, and each entry contains a tuple of payoffs to the row and column players, respectively. For example:
\begin{equation}
    \bbordermatrix{
        ~ & L & R \cr
        U & (a\row, a\col) & (b\row, b\col) \cr
        D & (c\row, c\col) & (d\row, d\col) \cr}.
\end{equation}

We separate this table into two matrices: the row payoff matrix $\bm{A}$, containing the first entries of each tuple; and the column payoff matrix $\bm{B}$, containing the second entries. A mixed strategy for the row (column) player is a probability vector whose dimension equals the number of rows (columns) of the payoff table. If the row player adopts strategy $\bm{p}$, her possible payoffs against a pure strategy playing column player is the portfolio $\bm{a}=(\bm{A}^T)\bm{p}$. Similarly, if the column player adopts strategy $\bm{q}$, her portfolio is $\bm{b}=\bm{Bq}$.

Players generally aim to maximize their payoffs, typically assuming that their opponents are rational. When this assumption fails (e.g.,\ when opponents behave maliciously), a player's degree of risk aversion becomes a key factor in strategy selection. Risk aversion can be measured in a variety of ways \cite{meyer06,pratt64}. Here, we measure the risk aversion of a player's strategy when facing an opponent whose behaviour cannot be trusted to be fully rational in the following way. Consider a column player who is unsure of the rationality of their opponent (or indeed if they are malicious). The worst payoff to this player over all possible outcomes is $\min(\bm{B})$. Let $v\col$ denote their maximin value --- their highest guaranteed payoff against any opponent. When $v\col > \min(\bm{B})$, we define the column player's risk aversion for employing strategy $\bm{q}$ as:
\begin{equation}
    r\col(\bm{b}, \bm{B}) = \frac{\min(\bm{b}) - \min(\bm{B})}{v\col  - \min(\bm{B})}, \label{eqn:risk}
\end{equation}
where $\bm{b}$ is the portfolio for strategy $\bm{q}$. This equation maps $\bm{b}$ to a real value, rescaling the payoff of the portfolio between the worst-case payoff ($r\col=0$) and the maximin guarantee ($r\col=1$) with intermediate values corresponding to partial risk aversion. We characterize a player by specifying a threshold value of her risk aversion $\theta\col$, where she will restrict herself to strategies with risk aversion that meet or exceed it. Note that this measure is meaningful only when $\min(\bm{B}) < v\col$, since no improvement over the worst outcome is possible otherwise.

For our model of partially malicious behavior, the objective of a partially malicious player is twofold. They prioritize obtaining a payoff that satisfies their basic needs, a minimum value they require. Once that payoff is secured, they act maliciously by minimizing their opponent’s payoff. We proceed by assuming that the column player is partially malicious. Such a player's decision making can be determined by the following function that takes both players' payoffs along with a specified minimum payoff threshold and map them to a value:
\begin{equation}
    U(\pi\row, \pi\col, \phi\col) = \begin{cases}
        -\infty, \text{ if } \pi\col < \phi\col, \\
        g(\pi\row), \text{ if } \pi\col \geq \phi\col.
    \end{cases}
\end{equation}
$\phi\col$ is the minimum payoff the column player is satisfied with, and $\pi\row$ and $\pi\col$ are the payoffs of the row and column player, respectively. The function $g(\pi)$ is strictly decreasing in $\pi\row$, ensuring malicious behaviour. For simplicity, we set $g(\pi) = -\pi$. We emphasize that, although this resembles a typical utility function, it does not satisfy the continuity axiom of expected utility maximization \cite{neumann44}. The negative infinity simply reflects the player's priority of achieving the minimum required payoff before optimizing their malicious objective. In practice, a sufficiently low real value, smaller than the minimum value of $g$ in the game, can be used to maintain consistency with the framework of expected utility.

\subsection{Truncation selection}

Independent truncation selection models the evolution of the state of a population with a mixture of player who play pure strategies of a symmetric game \cite{morsky16}. The population state $\bm{x}$ is a vector whose entries are the proportions of the population of each strategy. The evolutionary process begins by specifying a selection pressure, a fixed threshold payoff for survival. Players compete against one another, earning payoffs from each encounter with an opponent. This process can be implemented by simulating round-robin tournaments with stochastic payoffs \cite{fogel97,fogel11}, which generates a distribution of the payoffs averaged over the competitions. Payoffs may also be represented by distributions and studied with continuous dynamical systems \cite{morsky16}. Regardless of how payoffs are particularly earned, at the end of the competition, players with averaged payoffs below the threshold are culled and do not reproduce. The remaining players each reproduce at the same rate so as to replace the culled players. In this way, players need to only have adequate payoffs that meet the threshold, and receive no further benefit for exceeding it.

Our model of truncation selection assumes a large number of interactions between players. Thus, the payoff variance is negligible and players effectively earn their fitnesses (i.e., their payoffs averaged over all interactions), the entries of the row payoff matrix $\bm{A}$ (see Appendix \ref{app:truncation_selection} for further details). In our agent-based simulations, we initialize a population of $N$ players of randomly determined pure strategies. In each round of the game, players earn the average payoff from each interaction and those with payoffs below the threshold $\phi$ are culled. The remaining population reproduces to maintain the population size, rounding to the nearest whole number. This evolution is iterated until equilibria or extinction. Note that one round may be all that is required to reach an equilibrium.

Approaching this problem analytically, note that if the culling threshold payoff $\phi$ is less than or equal to the column maximin value of $\bm{A}$, then there exists a set of states where all player types meet or exceed the threshold. Thus, the population remains unchanged, and thus we treat such states as equilibria. We call the set of such equilibria the safe space. The set of equilibria for a given support typically form a convex subset of the strategy simplex, and the overall safe space is the union of safe spaces for each support (note that the union of equilibria across supports is not convex in general). Assuming a full-support state (for other supports, one simply cross out the corresponding rows and columns that are not in support), we can compute these equilibria via the following polyhedral conversion problem: 
 \begin{equation}
 \begin{aligned}
 \text{convert from H- to V-representation: } \\
     1 = \bm{1}^T\bm{x} \\
    \bm{0} \leq \bm{Ix} \\
    \phi\bm{1} \leq \bm{Ax}
 \end{aligned}
\end{equation}
The first two constraints restrict the solution space inside the probability simplex, while the last enforces the equilibrium condition (i.e.,\ that the payoffs of all player-types meet or exceed $\phi$).

When $\phi$ is less than or equal to the column maximin value of $\bm{A}$, the vertices in V-representation form a convex subset of the probability simplex. When the $\phi$ is exactly the column maximin value of $\bm{A}$, the V-representation is exactly the set of maximin strategies (typically single points, but can be a convex set in general). When $\phi$ is above the column maximin value of $\bm{A}$, the polyhedron is the empty set. Thus, all states will have at least one player type go extinct, reducing the game to a sub-game. States outside of a non-empty safe space may evolve toward this space, toward a reduced game, or toward complete extinction of all player types.

\section{Results}

\subsection{Games with malicious players}

We begin with a scenario where the column player is rational and the row player is partially malicious, given a specified risk aversion threshold $\theta\row \in (0,1)$. Knowing that her opponent is malicious, it would be intuitive for the column player to adopt a maximin strategy as a defensive measure. However, due to the minimum payoff requirement, the row player's strategy space may be restricted. So, even if the column player expose herself to risk by playing a strategy not in the set of maximin strategies, the row player might be unable to fully exploit the worst case. Since, the column player can threaten the loss of his minimum requirement. This creates the potential for the column player to secure a payoff at least as high as the maximin value through a computation method that leverages the row player's restricted strategy space.

To begin the analysis, consider the row player's strategy space as a convex hull of probability vectors. The vertices of this space can be obtained by solving a polyhedral representation conversion problem (see Appendix \ref{app:PRC_methods} for a review): 
\begin{equation}
 \begin{aligned}
 &\text{convert from H- to V-representation: } \\
     &1 = \bm{1}^T\bm{x} \\
    &\bm{0} \leq \bm{Ix} \\
    &\phi\bm{1} \leq \bm{Ax}
 \end{aligned}
\end{equation}
This polyhedral conversion problem yields the vertices of the restricted strategy space for the malicious player. Let these resulting vertices (strategies) be $\bm{p}_0, \bm{p}_1, ... ,\bm{p}_k$. By the fundamental idea of linear programming, one of these strategies yields the worst payoff the row player can make the column player accept. Therefore, the column player's payoff is $\pi\col = \min(\bm{p}_0^T\bm{Bq}, \bm{p}_1^T\bm{Bq}, \ldots, \bm{p}_k^T\bm{Bq})$ with $\bm{q}$ denoting her strategy.

Selecting column player's optimal strategy in this scenario is equivalent to solving a generalized maximin problem. To see this, define $\bm{P}_\phi=[\bm{p}_0, \bm{p}_1, \ldots ,\bm{p}_k]$. Then, the following maximin problem is equivalent to maximizing $\pi\col$:
\begin{equation}
\begin{aligned}
    \text{maximize: }& v\col \\
    \text{subject to: }& \bm{P}_\phi^T\bm{Bq} \geq v\col \bm{1} \text{ and } \bm{1}^T \bm{q} = 1, \\
    &\bm{q} \geq 0.
\end{aligned}
\end{equation}
Each entry of $\bm{P}_\phi^T\bm{Bq}$ is the column player's payoff against a row player's vertex strategy. By maximizing the minimum value, we obtain a defensive strategy robust against the row player's restricted strategy set. One can see that this linear program as a generalization of maximin problems, where our opponent's mixed strategies are limited in the convex hull of columns of the $\bm{P}_\phi$ matrix. When the row player is unrestricted in his strategy, $\bm{P}$ reduces to the identity matrix (since all pure strategies are available), and this problem recovers the maximin strategy. Depending on how severely row player's strategy space is restricted by his minimum requirement, this optimization may or may not yield a different strategy for the column player, but her payoff will always be at least as high as the classical maximin value.

We end the analysis by giving a numerical example for the game
\begin{equation}
    G=(\bm{A},\bm{B}) =
        \begin{bmatrix}
        (-62, 13) & (44, -33) & (62, -63) \\
        (-42, -76) & (4, -90) & (62,34)  \\
        (24, -30) & (-77, -63) & (-68, -39) \\
        (28, 85) & (80, -33) & (53, -24)
    \end{bmatrix}, \label{eq:exampleGame}
\end{equation}
where the row player is partially malicious with a risk aversion threshold of $0.22$ (and thus his minimum requirement is approximately $-53.89$), and the column player is rational. A maximin strategy for the column player is $[0.61, 0, 0.39]^T$ and the corresponding worst-case payoff is $-33.48$. However, since the row player's strategy set is restricted, the generalized maximin procedure yields an enhanced column strategy $\bm{q} \approx [0.54, 0, 0.46]^T$ and the worst payoff the opponent can enforce is $\pi\col \approx -31.73$.

Now consider how the row player computes his strategy. When he plays strategy $\bm{p}$, the possible payoffs available to the column player are $\bm{B}^T\bm{p}$. Since the column player is rational, she will choose her best response to these, so his goal is to minimize her best possible payoff while satisfying his own minimum payoff constraint, which can be represented by the following optimization problem:
\begin{equation}
\begin{aligned}
    \text{minimize: }& v\col \\
    \text{subject to: }& \bm{B}^T\bm{p} \leq v\col \bm{1},\text{and } \bm{A}^T\bm{p} \geq \phi\row\bm{1},\text{ and } \bm{1}^T \bm{p} = 1, \\
    &\bm{p} \geq 0.
\end{aligned}
\end{equation}
This is a generalized minimax problem, which extends the standard formulation with additional constraints from the minimum payoff requirement. Solving this program yields a strategy that simultaneously satisfies this constraint and minimizes the column player's best response payoff. This upper bound is theoretically the same value as the rational player's enhanced payoff derived earlier, agreeing with the famous minimax theorem \cite{simons95}. For the row player in the example game above, a generalized minimax strategy is approximately $[0.10, 0.13, 0.77,0]^T$, and the corresponding best possible payoff for the column player under this strategy is again $-31.73$, which is in agreement with the minimax theorem.

\subsection{Independent truncation selection}

\begin{figure}[!ht]
    \centering
    \includegraphics[width=0.7\textwidth]{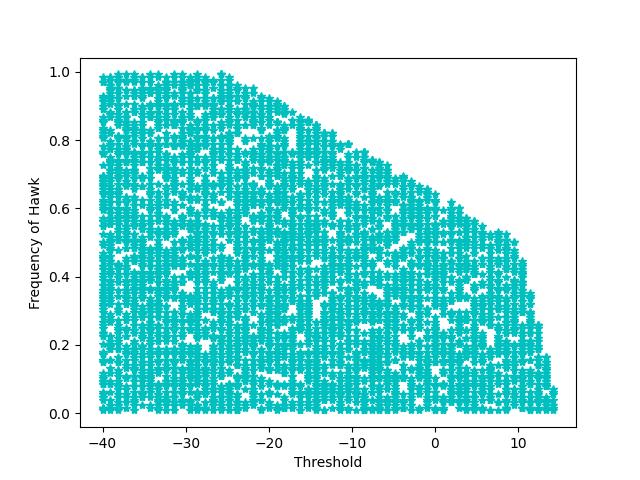}
    \caption{Equilibrium frequency of hawks under truncation selection vs.\ the payoff threshold $\phi$. The upper bounds of the outcomes have two distinct linear boundaries with a turning point at $\det(\bm{A}')=0$.}
    \label{fig:truncation}
\end{figure}

We begin by analyzing results of an agent-based numerical simulation of a Hawk-Dove game, using a similar but simplified setup as \cite{fogel97,fogel98,morsky16}. Assuming a large number of competitions between players, we use expected payoffs instead of stochastic ones, which are the fitnesses; implications of this modeling choice are discussed in more depth in Appendix \ref{app:truncation_selection}. Payoffs for the two types of players, hawks and doves, are determined by the row payoff matrix:
\begin{equation}
    \bm{A} =
        \begin{bmatrix}
        -25 & 45 \\
        5 & 15
    \end{bmatrix} \label{eq:hawk-dove}
\end{equation}
with the first strategy representing hawks and the second doves. Figure \ref{fig:truncation} shows the equilibrium frequency of hawks as a function of $\phi$. Consider the shape of the distribution of data samples. The upper bounds of the data have two distinct linear boundaries separated by a turning point. To explain this observation, consider the payoff matrix shifted by the payoff threshold: $\bm{A}' = [a_{ij} - \phi]$. If a population state $\bm{x}$ satisfies $\bm{A}'\bm{x} \geq \bm{0}$, then both hawks and doves will have payoffs above $\phi$. This is connected to computing the column maximin value of $\bm{A}$, which is $15$. Thus, for all $\phi>15$, there exists no state $\bm{x}$ such that hawks and doves both have fitnesses above $\phi$. Conversely, for $\phi<-25$, the fitnesses of both hawks and doves will be great than $\phi$ for every state $\bm{x}$.

Although extending the horizontal axis of Figure \ref{fig:truncation} beyond $[-25,15]$ is unnecessary for this model, it can be necessary when there is significant variance in payoffs, as in the models of \cite{fogel97,fogel98,morsky16}. Because, outliers are possible, and thus the boundary is different. For $\phi<-25$, culling and extinction are rare, but not impossible for stochastic payoffs. For $\phi>15$, primarily populations with states equal to a column maximin strategy (which is exactly $\bm{e}_2$ in this case) can avoid extinction. Increasing $\phi$ shrinks this range linearly into a single point. The turning point is when $\det(\bm{A}')=0$, which is at the threshold $\phi=10$. We care about the determinant because it would be possible for non-trivial states to exist, making the fitnesses of all players meet the threshold. For $\phi<10$, we can explain the linear boundary by computing the range of safe spaces where we expect no extinctions. This is relatively easy for $2\times2$ games, because the set of safe spaces is a subset of the probability hyperplane with dimension $2$ (a line segment). The comparison to the original result in \cite{morsky16} is included in Appendix \ref{app:truncation_selection} where we show that the difference between our result and theirs is due to low population sizes and stochastic payoffs with non-negligible variances. As the population size increases, their results converge to ours. 

We can obtain the convex hull of vertices by finding the two extreme points of the line segment using the following linear program (note that for higher dimensions, this must be converted to a polyhedral representation conversion problem):
\begin{equation}
\begin{aligned}
    \text{minimize: }& \bm{e}_1^T\bm{x} \\
    \text{subject to: }& \bm{A}'\bm{x} \geq 0 \text{ and } \bm{1}^T \bm{x} = 1, \\
    &\bm{x} \geq 0.
\end{aligned}
\end{equation}
The linear objective function is just a vector in the standard basis of $\mathbb{R}^2$, meaning that we are minimizing an entry in $\bm{x}$. Doing this for both entries allows us to obtain the convex hull. When $\phi<10$, the extreme states tell us that hawks become extinct before (since the first row of $\bm{A}'\bm{x}$ is zero). When $\phi>10$, doves go extinct before hawks for a similar reason (the second row of $\bm{A}'\bm{x}$ is zero). The two slopes in Figure \ref{fig:truncation} are given by the following two relations:
\begin{equation}
    \left(\begin{bmatrix}
    -25 & 45 \\
    5 & 15  
    \end{bmatrix} 
    -\phi\begin{bmatrix}
    1 & 1 \\
    1 & 1  
    \end{bmatrix}\right)
    \begin{bmatrix}
    x_1 \\
    x_2
    \end{bmatrix}
    = \begin{dcases}
        \begin{bmatrix} 0 \\ \pi_2 \end{bmatrix}, \text{ if } \phi < 10, \\
        \begin{bmatrix} \pi_1 \\ 0 \end{bmatrix}, \text{ if } \phi > 10, \\
    \end{dcases}
\end{equation}
where the unspecified fitnesses (denoted by $\pi_1$ and $\pi_2$ for hawks and doves, respectively) are non-negative. Solving these equations yields the slopes of the two upper boundaries: $-1/70$ and $-1/10$. 

By the above arguments, we conclude that no culling occurs for $\phi$ below the minimum possible payoff, and only rare players survive for $\phi$ above the column maximin value. This analysis provides a direct way to compute a naive safe space, assuming full support (about which we will elaborate below).

For higher dimensional games, the computation of safe spaces is more complex. It is possible that a player type goes extinct, irreversibly reducing the game to a lower-dimensional sub-game for which there may be a different safe space. We adopt the game theoretic notion of a strategy being in the support to express the surviving strategies in such a scenario. Then, given a specified threshold payoff, we compute safe spaces for all possible supports and take the union to find the set of all safe strategies.

\begin{figure}[!ht]
    \centering
        \includegraphics[width=0.7\textwidth]{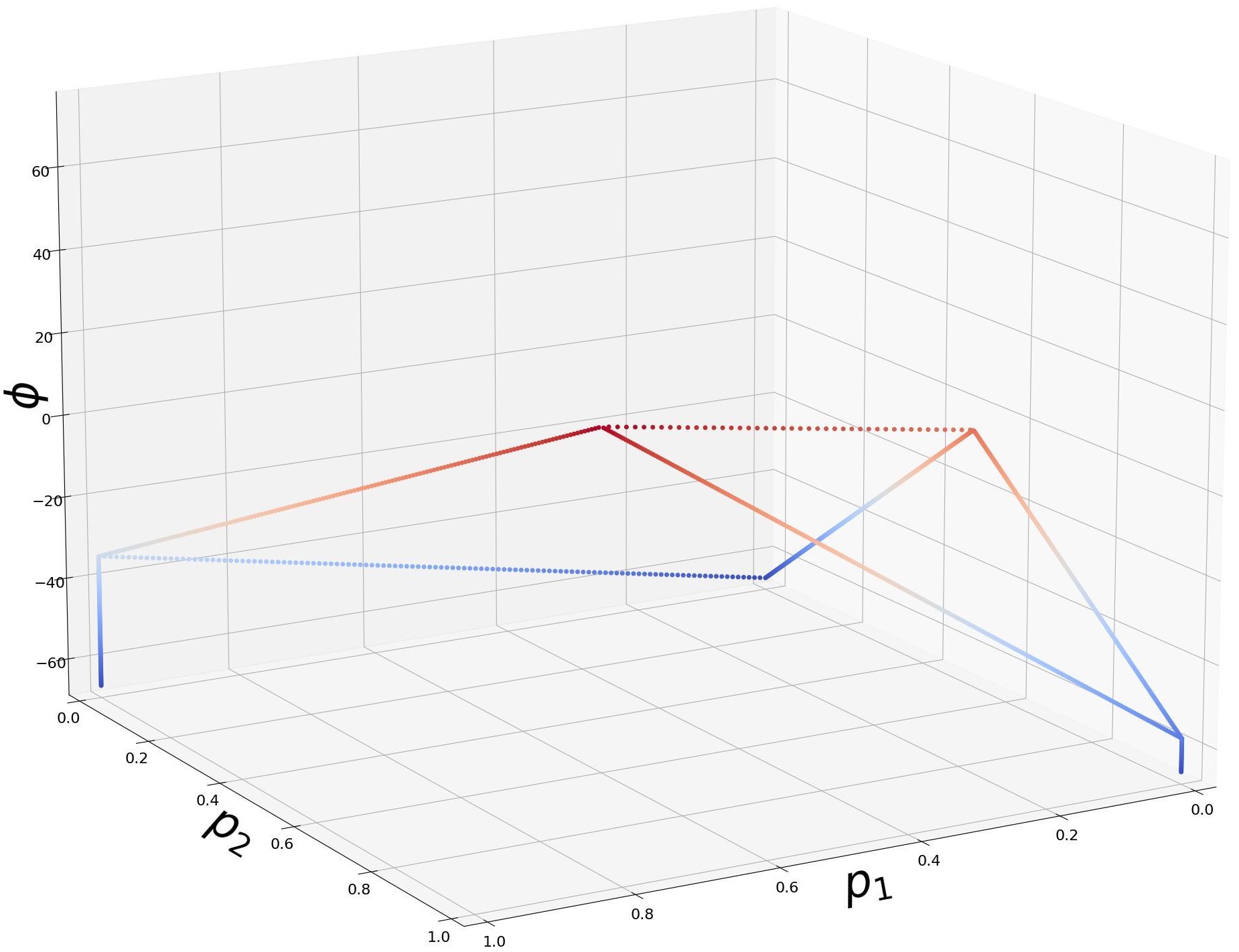}
    \caption{Visualization of the full support safe space for a $3\times3$ game. Each horizontal slice of the shape is the polyhedron corresponding to the safe space given by the threshold $\phi$, where a colder colour represents a lower threshold. Vertices of the polyhedra vary linearly and converge to maximin strategies (typically unique, but not unique in general) at the maximin value.} \label{fig:independent3D_111}
\end{figure}

\begin{figure}[!ht]
    \centering
        \includegraphics[width=0.7\textwidth]{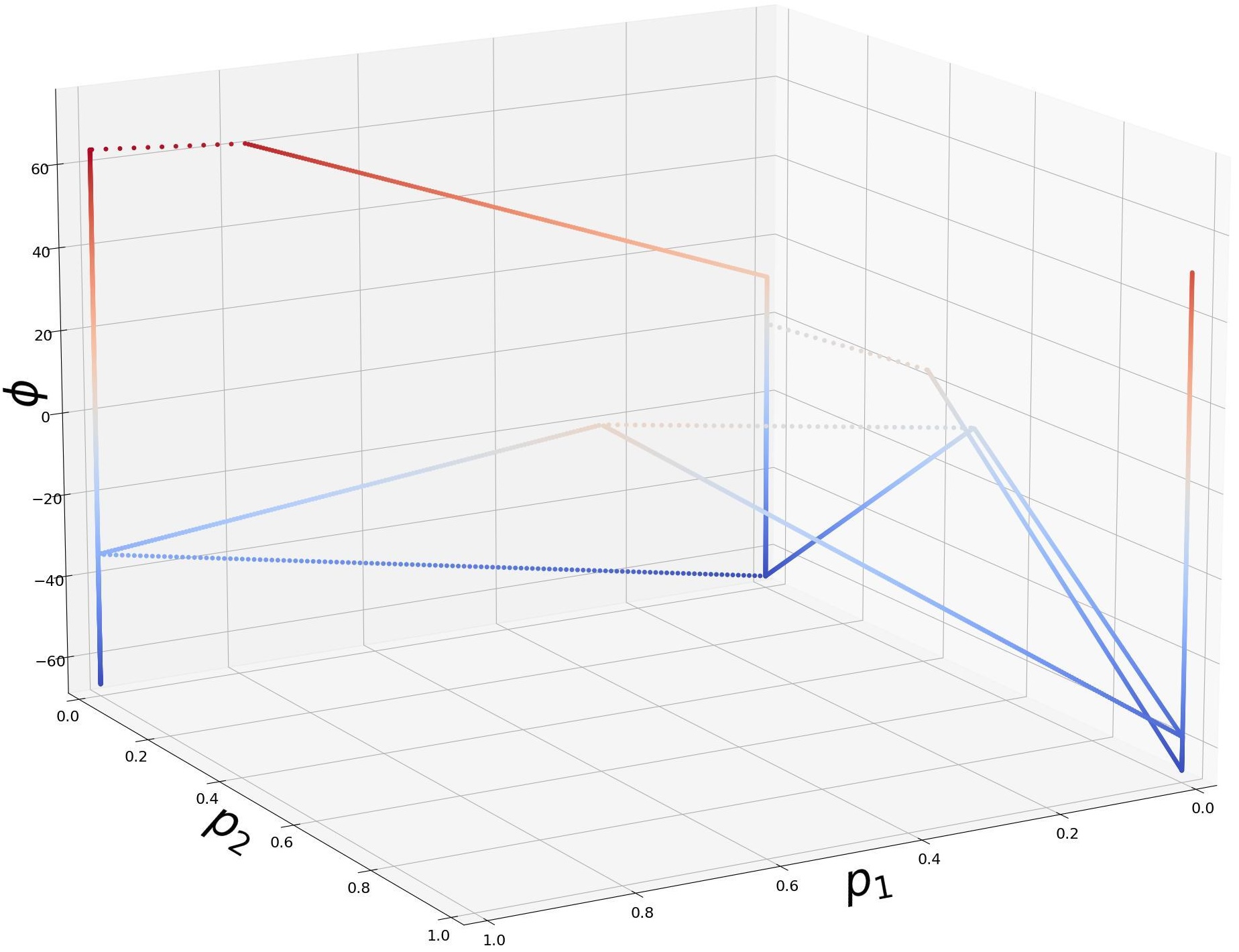}
    \caption{Visualization of the overall safe space for a $3\times3$ game. This shape is a union of convex subsets (though the union is not convex in general) for all sub-games obtained by using all possible supports.} \label{fig:independent3D_allSupport}
\end{figure}

To see this procedure, consider a $3 \times 3$ game with the following payoff matrix:
    \begin{equation}
    \bm{A} =
        \begin{bmatrix}
        64 & -58 & 50 \\
        -34 & 51 & -66 \\
        75 & -1 & 11
    \end{bmatrix}. \label{eq:3x3game}
\end{equation}
This game has the pure Nash equilibria $(\bm{e}_1,\bm{e}_3)$, $(\bm{e}_3,\bm{e}_1)$, and $(\bm{e}_2,\bm{e}_2)$, and the mixed Nash equilibria $(\bm{p}_1,\bm{q}_1)$ and $(\bm{p}_2,\bm{q}_2)$ where $\bm{p}_1 = \bm{q}_1 =[0, 77/129, 52/129]^T$ and $\bm{p}_2 = \bm{q}_2 =[39/50, 0, 11/50]^T$. And it has the maximin value of approximately $6.24$. The numerical results for the safe spaces of this game are generated by the following procedure and plotted in Figure \ref{fig:independent3D_111} and \ref{fig:independent3D_allSupport}, where the first plot is the safe space only considering the full support, and the second plot is the safe space where all supports are considered. Since the vertices of the safe spaces are three dimensional probability vectors, there are only two degree of freedom. Thus, we only need to choose two components to record and the remaining axis to record the thresholds. In our plot, the $x$-axis is the first coordinates of the vertices, the $y$-axis is the second, and the $z$-axis is the axis of thresholds. For different threshold values, we can compute the vertices of the safe spaces. This can be done by solving the conversion problem stated earlier, and then recording the values for each vertex and plot the three axes.

Given a payoff matrix $\bm{A}$, our results are generated in the following way. First, we compute the column maximin value of the row payoff matrix. Since, the vector of fitnesses for each player type are convex combinations of the columns of the row payoff matrix where the coefficients are the frequencies of each player type. We are looking for the maximum possible value of the minimum entry in the fitness vector under all possible convex column combinations. This is exactly the column maximin value. This maximin value is the maximum threshold given that a full support safe space exists. Note that the maximin value may vary if the support is changed due to the irreversibility in truncation selection. Following this, we generate a set of thresholds to observe ranging from the minimum of the payoff matrix to its maximum. When a threshold is less than or equal to the minimum, no culling happens, and when greater than the maximin value, no non-empty safe space exists for full support with respect to the maximin value. 

\section{Discussion}

We have applied the concept of safe spaces, linear programming, and polyhedral representation conversion to two subjects of game theory: partially malicious players and independent truncation selection. Our aim was to develop methods that allow players to manage worst-case scenarios efficiently.

Regarding partially malicious players, we constructed a simple model for players who are partially malicious. Such players only become malicious after securing a specified minimum payoff. If we require defensive plays against such players, we may use a generalized maximin procedure. In the case where both players have set their risk thresholds sufficiently low, both will have non-empty safe strategy sets and thus the game could be formulated as a bilinear game \cite{garg11}. Future studies could develop refine our decision-rule, and the impetus for maliciousness could be directly modeled such as by including relational utility. 

With respect to truncation selection, we showed that the equilibria are convex subsets of the probability space for each support. As the selection threshold varies linearly, the vertices of safe spaces also change linearly. When the threshold is exactly equal to the maximin value, the safe space converges to the set of maximin strategies. When the threshold is above the maximin value, extinction happens, and the game is irreversibly reduced. In general, the states do not converge to one of the Nash equilibria of the underlying game. These findings hold under the assumption of large populations, though prior work has shown how small populations can exhibit unexpected behaviours \cite{fogel98,morsky16}. Using the polyhedral conversion approach, we can compute the set of safe spaces analytically or less computationally intensively(with respect to the population size) than with stochastic simulations. Note that polyhedral conversion algorithm can still become complex when the dimension is large, so one might choose from conducting the actual simulation, or analyzing the vertices.

More broadly, our paper provides two additional scenarios in game theory requiring dense computations that benefit from linear programming and polyhedral conversion. This reinforces that value of expressing game theoretic concepts in linear algebraic language, and converting problems into feasible problems that these algorithms can solve.

\subsection*{Code and Data Availability}
Code is available at github.com/bmorsky/LP\_malice\_and\_truncation.

\appendix

\section{Appendix}

\subsection{Linear programming methods for computing maximin values} \label{app:LP_methods}

The maximin value for a player is the maximum possible value she can be guaranteed to obtain, and the strategies guaranteeing this value are called maximin strategies. In this section, we quickly review a linear programming method to compute, without loss of generality, the column player's maximin strategy of a game.
    
One of the column player's maximin strategies is a convex combination of the columns of the payoff matrix $\bm{B}$ where the portfolio vector $\bm{b} = \bm{Bq}$ has its minimum entries maximized over the convex hull of the columns of $\bm{B}$. The value of the minimum entries, $v\col$, is called the maximin value of this game for the column player. One can see that adding a scalar to all entries of $\bm{B}$ will not change the column player's maximin strategies: the maximin value of $\bm{B}$ is just shifted by this scalar. One can also see that any $\bm{B}$ with a positive column will have a positive maximin value, because the value is at least the minimum entry of the pure strategy corresponding to this column. The method for finding a maximin strategy for the column player can be stated as the following optimization problem:
\begin{equation}
\begin{aligned}
    \text{maximize: }& v\col \\
    \text{subject to: }& \bm{Bq} \geq v\col \bm{1} \text{ and } \bm{1}^T \bm{q} = 1, \\
    &\bm{q} \geq 0.
\end{aligned}
\end{equation}
Note that we do not know $v_{col}$ when we construct the linear program. To proceed, we can convert this optimization problem into a linear programming problem by expressing the objective function as a linear function of the variable vector.

An easy way to do this conversion is by first adding a scalar $k$ to every entry of $\bm{B}$ such that $v\col > 0$, one can arbitrarily choose a column to make it positive. Then, rather than using a probability vector $\bm{q}$, we use a non-negative vector $\bm{x}$. Dividing both sides of the inequality constraints by $v\col$, setting $\bm{x} = \bm{q}/v\col$, and ignoring the constraint $\bm{1}^T\bm{q} = 1$ for the moment, we obtain the following optimization problem:
    \begin{equation}
    \begin{aligned}
        \text{maximize: }& v\col \\
        \text{subject to: }& \bm{Bx} \geq \bm{1}, \\
        &\bm{x} \geq \bm{0}.
    \end{aligned}
    \end{equation}

Because $\sum_i q_i=1 \implies \sum_i x_i = 1/v\col$, we can express the objective function as a linear function of $\bm{x}$, namely $1/v\col = \bm{1}^T\bm{x}$. Now, the optimization problem can be seen as a linear programming problem:
    \begin{equation}
    \begin{aligned}
        \text{minimize: }& \bm{1}^T \bm{x} \\
        \text{subject to: }& \bm{Bx} \geq \bm{1}, \\
        &\bm{x} \geq \bm{0}
    \end{aligned}
    \end{equation}
The solution for this problem yields a particular maximin strategy $\bm{q} = \bm{x}/\sum_i x_i$. The column maximin value of $\bm{B}$ is the reciprocal of $\bm{1}^T \bm{x}$, shifted back by the scalar $k$ we have previously added: $v\col = (1/\sum_i x_i)-k$.

Linear programming can be seen as iterating over the vertices of the polytope specified by the constraints (if feasible). The one that maximizes the linear objective function is a particular solution, based on the observation that solutions of such problems will always have intersection with at least one vertex of the polytope.

\subsection{Polyhedral representation conversion} \label{app:PRC_methods}

Since it is useful to use tools for solving polyhedral representation conversion problems when studying truncation selection, here we will briefly review this topic but also direct the reader to \cite{avis02} for further details. Again, without loss of generality, let's assume we are the column player. Then the portfolio vectors for each pure strategies are the columns of $\bm{B}$. The set of possible column portfolios for this game would be the convex hull of the columns, since mixed strategies are convex combinations over the columns. This set can be seen as a convex polyhedron in $\mathbb{R}^n$, where $n$ is the number of rows of $\bm{B}$. A convex polyhedron has two equivalent expressions, namely V-representations and H-representations.

Under V-representations, the collection of the vertices of the convex hull is used to define the unique convex polyhedron. The set of possible portfolio vectors are naturally given in V-representations. To be specific, say $\bm{b}_0$,  $\bm{b}_1$, \ldots, $\bm{b}_m$ are the columns of $\bm{B}$. These vectors correspond to the portfolio vectors of the player's pure strategies. Then, the V-representation of the convex polyhedron equivalent to the set of possible portfolio vectors is $\conv(\bm{b}_0,  \bm{b}_1, \ldots, \bm{b}_m)$, where $\conv(\cdot)$ stands for convex hulls.

Under H-representation, the polyhedron is described by a set of linear half-spaces, the intersection of which is exactly the convex polyhedron. A set of linear inequalities and equalities are typically used and arranged such that they are in form $\bm{y}_i \leq \bm{C}_i\bm{x}$ and $\bm{y}_e = \bm{C}_e\bm{x}$. Here $\bm{C}_i$ and $\bm{C}_e$ are the matrices consisting of the coefficients of the inequalities and equalities, respectively. 
    
In this paper we avoid repeatedly emphasizing the convex property by assuming a polyhedron is convex unless specified otherwise. The problem of converting a polyhedron from one representation to another is called a polyhedron conversion problem. Given feasibility, there are computational algorithms \cite{avis91, fukuda95} that solve this type of problem (though they can be computationally expensive for high dimensions). 

\subsection{Observing the vanishing variance for the Hawk-Dove game} \label{app:truncation_selection}

\begin{figure}[!ht]
    \centering
    \centering
     \begin{subfigure}[]{0.45\textwidth}
        \includegraphics[width=\textwidth]{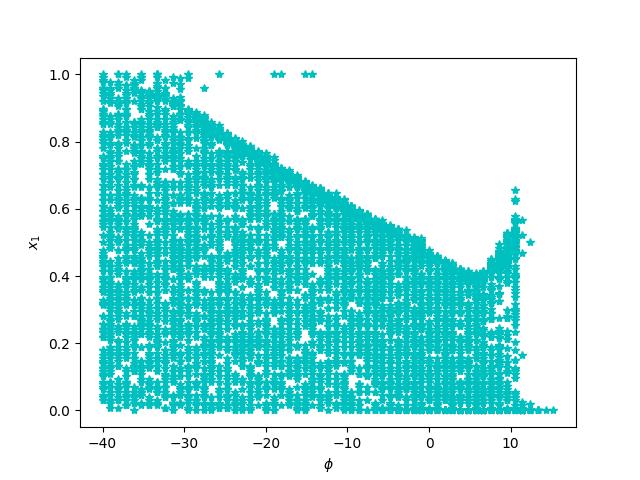}
        \caption{population size = 128}
     \end{subfigure}
     \begin{subfigure}[]{0.45\textwidth}
        \includegraphics[width=\textwidth]{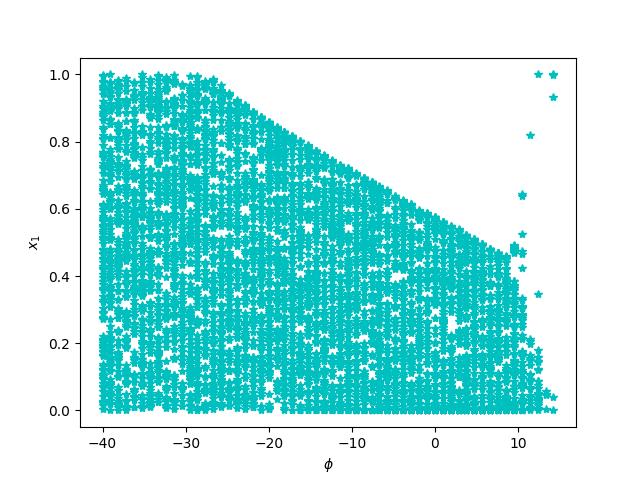}
        \caption{population size = 1024}
     \end{subfigure}
     \begin{subfigure}[]{0.45\textwidth}
        \includegraphics[width=\textwidth]{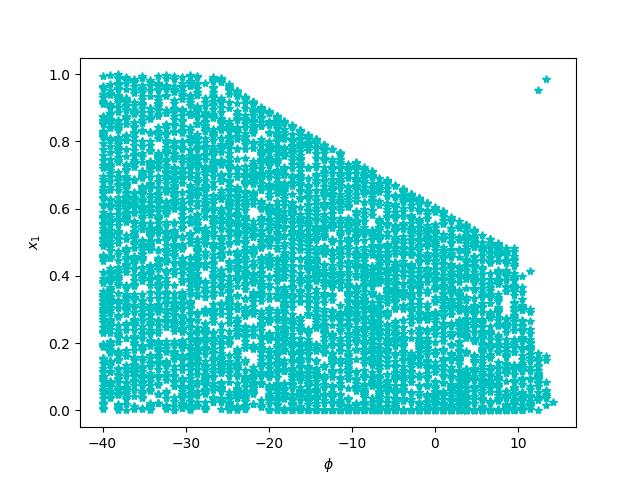}
        \caption{population size = 4096}
     \end{subfigure}
    \caption{Plotted here are the results of typical runs of the model in \cite{morsky16} with a population sizes $N=128,1024,4096$.}
    \label{fig:stochastic_payoffs}
\end{figure}

Here we elaborate on the dynamics of the Hawk-Dove game under independent truncation selection explored in \cite{morsky16}. To begin, we reproduce their original agent-based simulation result in Figure \ref{fig:stochastic_payoffs} to contrast with Figure \ref{fig:truncation} from the main text. Before explaining these differences, we specify the truncation algorithm.

Specify a population size $N$ and a survival threshold $\phi$. Then, randomly determine the initial frequency of hawks $x_1$ noting that the frequency of doves is $x_2$. Then, for each hawk and dove, there is a round robin tournament (which has high computational complexity in population size). We can replace the total mixing by expressing the fitness for interactions between hawks and doves as stochastic variables $X_{ij} \sim \mathcal{N}(\mu_{ij}, \sigma_{ij}^2)$ where $i$ is the strategy earning the payoff and $j$ is their opponent. $X_1 \sim \mathcal{N}(\mu_h, \sigma_1^2)$ and $X_2 \sim \mathcal{N}(\mu_2, \sigma_2^2)$ are the stochastic variables of the average payoffs of hawks and doves, respectively, which are a mixture of the other distributions. To obtain the expressions of means and variances of these distributions, we use the fact that for an individual hawk, it has interacted with approximately $x_1N$ hawks and $x_2N$ doves for fairly large populations. Then the cumulative payoff is averaged by population size, so we have the following:
\begin{align}
    X_1 &= \frac{1}{N}\left(\sum_{1}^{x_1N}{\mathcal{N}(-25, 75^2)}\right) + \frac{1}{N}\left(\sum_{1}^{x_2N}{\mathcal{N}(45, 15^2)}\right) \notag \\
        &= \frac{x_1}{x_1N}\left(\sum_{1}^{x_1N}{\mathcal{N}(-25, 75^2)}\right) + \frac{x_2}{x_2N}\left(\sum_{1}^{x_2N}{\mathcal{N}(45, 15^2)}\right) \notag \\
        &= x_1\mathcal{N}\left(-25, \frac{75^2}{x_1N}\right)+ x_2\mathcal{N}\left(45, \frac{15^2}{x_2N}\right).
\end{align}
By the calculation rules for linear combinations of normal random variables, we are able to obtain $\mu_1 = -25x_1 + 45x_2 = -70x_1 + 45$ and $\sigma_1 = \sqrt{ 75^2x_1/N + 15^2x_2/N}=\sqrt{5400x_1+225}/\sqrt{N}$. 
Similarly for doves, we have $\mu_2 = 5x_1 + 15x_2 = -10x_1 + 15$ and $\sigma_2=\sqrt{(15^2x_1/N + 25^2x_2/N}=\sqrt{-400x_1+625}/\sqrt{N}$. An immediate observation is that the standard deviations are inversely dependent to the root of population size. As $N \to \infty$ the variances go to zero, and we obtain the scenario in the main text. From our exploration of these simulations, any population size above roughly $10,000$ will be able to converge closely to the result in the main text. However the round robin tournament simulation becomes computationally expensive as population size increases.

Another difference between Figure \ref{fig:truncation} and Figure \ref{fig:stochastic_payoffs} is that the upper limit of hawks is higher in the former figure than the latter. Note that the mean payoffs for both hawks and doves decrease as $x_1$ increases. However, the standard deviation for hawks increases as $x_1$ increases, and the standard deviation for doves decreases as $x_1$ increases. Thus, for a high initial hawk frequency, hawks have a higher chance of being below a low threshold than doves. However, for large $\phi$, the probability of doves obtaining adequate payoffs becomes low. Since the standard deviation of hawks is much larger, hawks have a greater chance to obtain adequate payoffs even when the threshold is above the maximin value. Although a large standard deviation means a higher probability for extremely low payoffs, when analyzing whether a population goes extinct, we only care about the probability for survival. Thus, for a threshold that is significantly larger than the mean payoffs, the strategy with a larger payoff variance will have a higher chance of survival.

\bibliography{LP}
\bibliographystyle{plain}

\end{document}